\documentstyle[12pt,worldsci,epsf]{article}
\begin{document}
\hspace{11.8cm}HUB-EP-98/57\\[-1cm]
\title{THE MECHANISM OF QUARK CONFINEMENT}
\author{Gunnar S.\ Bali\footnote{Talk given at 3rd International
Conference on Quark Confinement and Hadron Spectrum
(Confinement III), Newport News, VA,
7-12 Jun 1998. 
}\\
{\em Institut f\"ur Physik, Humboldt Universit\"at, Invalidenstra\ss{}e 110,
10115 Berlin, Germany}}

\maketitle
\setlength{\baselineskip}{2.6ex}

\vspace{0.7cm}
\begin{abstract}
I summarise recent lattice results on the QCD confinement mechanism
in the maximally Abelian projection. 
\end{abstract}
\vspace{0.3cm}

\section{Introduction}
The phenomenology of strong interactions contains two fundamental
ingredients: asymptotic freedom and the confinement of colour charges.
The former requirement led to the invention of QCD.
A mathematically rigorous proof that QCD as the {\em microscopic} theory of
strong 
interactions indeed gives rise to the {\em macroscopic} property of
linear quark confinement as
indicated by Regg\'e trajectories and quarkonia spectra
is, after a quarter of a century, still lacking.
Meanwhile, lattice gauge theory simulations have provided convincing
numerical evidence for this conjecture.

The difficulty in deriving infra red properties of QCD illustrates that
something qualitatively new is happening: unlike in previously
existing elementary physical theories, it is not possible to reduce everything
down to two-body interactions but collective excitations
of quark and gluon states have to be accounted for. For the first time,
excitations of the vacuum that are considered to be fundamental
do not occur as initial or final states anymore. Therefore, {\em
understanding} confinement, in my opinion, is one of the most exciting
challenges of modern physics. New physical and mathematical techniques that
can successfully be applied to non-perturbative QCD might be required
for dealing with other strongly interacting theories or theories
with a non-trivial vacuum structure, in general. Vice versa
techniques developed in a different context might help to
prove QCD confinement and in solving QCD. A recent example is the
proof of confinement in  SUSY Yang-Mills
theories~\cite{wittenseiberg}. Of course QCD as we know it does
not obey super-symmetry but nonetheless
such activities point into a promising
direction. Just as QCD can serve as a guinea pig and development
centre for non-perturbative techniques, lattice simulations help probing the
validity range of effective low-energy models or in verifying
certain conjectures.

QCD in itself is sufficiently complicated to keep many physicists busy.
Solving QCD is still important and even more so since (almost) everyone
believes in it.
It is widely accepted that perturbative QCD (pQCD) successfully describes
high energy scattering processes. However, without understanding
non-perturbative aspects of QCD, even in this case, it is not
clear why pQCD works at all. Only hadrons, leptons and photons but no
quarks or gluons are observed in experiment. In order to fill the gap,
the formation of colour-neutral
hadronic jets from quarks and gluons has to be
modelled. Furthermore, it is commonly conjectured that, once the
fundamental scattering on the
quark and gluon level has taken place, further interactions can
be neglected. It is demanding to derive the ingredients of
fragmentation models directly from truly non-perturbative QCD,
as well as to verify the factorisation
hypothesis.

Many phenomenologically important questions are posed in low-energy QCD
that eagerly await an answer:
is the same set of fundamental parameters (QCD coupling and quark masses)
that describes for instance the hadron spectrum consistent with
high energy QCD or is there
place for new physics? Are all hadronic states correctly
classified by the na\"\i{}ve quark model or do glueballs, hybrid states
and molecules play a r\^o{}le?
At what temperatures/densities does the transition to
a quark-gluon plasma occur? What are the experimental signatures
of quark-gluon matter? Can we solve nuclear physics on the quark and
gluon level?  Clearly, complex systems like
iron nuclei are unlikely ever to be solved from {\em first principles} alone
but {\em modelling} and certain {\em approximations} will always be required.
Therefore, it is desirable to test model assumptions,
to gain control over
approximations and, eventually, to derive low-energy effective
Lagrangians from QCD. Lattice simulations are a very
promising tool for this purpose.

In the past decades, many explanations of the confinement mechanism
have been proposed, most of which share the feature that topological
excitations of the vacuum
play a major r\^ole. A list of these theories includes the
dual superconductor picture of confinement~\cite{hooft,hooft2},
the centre vortex
model~\cite{ambjorn}, the 
instanton liquid model~\cite{diakonov}, and the anti-ferromagnetic
vacuum~\cite{savvidy}.
All these interpretations have been explored in lattice studies.
The situation with respect to an anti-ferromagnetic vacuum is still somewhat
inconclusive~\cite{antif}. Instantons seem to be more
vital for chiral symmetry related properties than for
confinement~\cite{instanton}.
Depending on the picture, the excitations
giving rise to confinement are thought to be
magnetic monopoles, instantons, dyons,
centre vortices, etc.. I would like to stress that the
above ideas are not completely disjoint
and do not necessarily exclude each other. For instance,
all the above mentioned topological excitations are
found to be correlated with each other
in numerical as well as analytical studies.

I will restrict myself to the, at present, most popular
superconductor picture which is based on the concept of electro-magnetic
duality after an Abelian gauge projection.
Recently, the centre vortex model has been resuscitated too. In the
latter picture,
the centre group that is directly
related to the traditional order parameter of the de-confinement
phase transition in finite temperature pure gauge theories, the
Polyakov line, plays an essential r\^ole. Another striking
feature is that --- unlike monopole currents --- centre vortices
form two-dimensional objects, such that in {\em four} space-time dimensions,
a linking number between a Wegner-Wilson loop and centre vortices can
unambiguously be defined, providing a geometric interpretation of
the confinement mechanism~\cite{poli}.
Unfortunately, I have not enough space to review these exciting
developments. Therefore, I refer the interested reader to
a review by Greensite~\cite{greensite}.

Another obvious omission
are questions related to chiral properties of QCD.
Real world QCD is not only a theory of gluons but includes quarks
which means that the de-confinement phase transition will eventually
be replaced
by a chiral phase transition. In the phase with broken chiral symmetry
colour charges are still being
anti-screened and linear confinement approximately holds.
However, if the binding energy within a hadron exceeds a critical value,
the hadron will break up into two or more colour-neutral
parts (string breaking).

Based on different confinement pictures various effective models of infra red
QCD have been proposed in the past.
Three such examples are the Abelian Higgs model~\cite{suzuki,poli},
the stochastic vacuum model~\cite{dosch} and dual QCD~\cite{baker}.
As well as {\em understanding} the mechanism of confinement, it is
desirable to verify aspects of these models and eventually to derive
certain model parameters directly from QCD.

This article is organised as follows. I will start with a brief motivation
and introduction into lattice methods in Sec.~\ref{sec:lat}, before
explaining the concept of the Abelian projection in Sec.~\ref{sec:abel}. 
Subsequently, I will introduce some notations and review lattice results
in Sec.~\ref{sec:res} and conclude with a summary of
answered and open questions.

\section{Why Lattice?}
\label{sec:lat}

Lattice methods allow for a somewhat
{\em brute force} but {\em first principles}
numerical evaluation of expectation values
of a given quantum field
theory that is defined by an action $S$. The lattice volume
and spacing are limited due to finite computer speed and memory. 
Simulations are performed in Euclidean space
and an analytic continuation to Minkowski space of numerical results that have
been obtained on discrete points with finite precision is virtually
impossible. Therefore, not every physically meaningful
quantity can be calculated in a straight forward manner.
The obvious strength of lattice methods are hadron mass determinations.
Even if one is unimpressed by post-dictions of experimental values
that have been known with high precision for decades, within
an accuracy of only 5~\%,
such simulations allow one to fix fundamental standard model
parameters like quark masses and QCD running coupling in the
low-energy domain. Of course, plenty of other applications of phenomenological
importance exist.

Unfortunately, only the lowest lying one or
two radial excitations of a hadronic state are
accessible in practice. Lattice predictions
are restricted to rather simple systems too.
Even the deuterium is beyond the reach of present day super-computers.
Therefore, it is desirable to supplement lattice simulations by
analytical methods. The computer alone acts as a {\em black box}.
In order to understand
and interprete the output values and to predict their dependence on
the input parameters, some modelling is required.
Vice versa, the lattice itself is a strong tool to validate models and
approximations. Unlike in the {\em ``real''} world, we can vary
the quark masses
$m_q^i$,
the number of colours $N_c$,
the number of flavours $n_f$, the temperature, the volume,
the space-time dimension of our {\em femto}-universe and even the
boundary conditions in order to expose models to thorough
tests in many situations. Instead of indirectly and in a somewhat
uncontrolled fashion deriving parameter values from experiment we
can compute {\em custom designed} observables.
What is thought to work in the real world ought to work on the lattice too!
Moreover, many models rely on certain approximations.
Experimentalists cannot switch off
quark flavours but we can!

In a lattice simulation, hyper-cubic Euclidean space-time is
discretised on a box with, say, $L^4$ lattice points or sites, $x$.
Two adjacent points are connected by an oriented bond or link
$(x,\mu)$. The product of
four links, enclosing an elementary square, is a plaquette. Quarks are
living on sites, gauge fields on links and the plaquette
determines the curvature within the $SU(N_c)$
group manifold and corresponds to the 
field strength tensor. For simplicity, we assume an isotropic lattice with
equal lattice spacing $a$ in all directions. The lattice spacing provides
an ultra violet cut-off on the gluon momenta $q<\pi/a$ and regulates the
theory.

We simulate an action
$S(\beta,m_q^i)$ on the lattice, which contains the quark masses as well
as an inverse
QCD coupling $\beta=2N_c/g^2$ as free parameters. By varying $\beta$
(and $m_q^i$) the lattice spacing $a$ is changed. Asymptotic freedom
tells us that the ultra violet cut-off is removed as
$\beta\rightarrow\infty$. The physical dimension of $a$ is determined by
calculating a dimensionful
quantity on the lattice and associating it to its experimental
value. If the right theory is being simulated we should be able to
reproduce all experimental mass ratios in the continuum limit
$a\rightarrow 0$, such that it becomes irrelevant what experimental
quantity we have chosen to set the scale\footnote{In
the results presented here,
I set the scale (somewhat arbitrarily) by the value
$\sqrt{\sigma}=440$~MeV for the string tension in case of $SU(2)$
and by the value $r_0=0.5$~fm in case of $SU(3)$ chromodynamics.
$r_0$ is the distance
at which $r_0^2\left.dV(r)/dr\right|_{r_0}=1.65$ where $V(r)$ denotes
the potential between two static colour sources,
separated by a distance $r$.}.
In practice one does not get all the quark masses right, such that
there are always systematic uncertainties in $a$, due to the ambiguity
of the choice of this experimental input quantity.

The discretised lattice action is formulated in a manifestly gauge-invariant
way and approaches the continuum action with
$a\rightarrow 0$. One ideally extrapolates to this limit at fixed physical
lattice extent, $La=$~const.. Later on, the thermodynamic limit
$La\rightarrow\infty$ should be investigated. Due to the finiteness of
a computer these two extrapolations, as well as extrapolations
and interpolations
between results obtained at
different quark masses, are subject to systematic uncertainties that
have to be carefully estimated. For most hadronic processes, a
lattice spacing $a<0.1$~fm is considered to be {\em close} to the
continuum limit while an extent $La\approx 2$~fm is comfortably
large to accommodate ground state mesons and baryons.

\begin{figure}
  \begin{centering}
\epsfysize=8truecm\centerline{\epsfbox{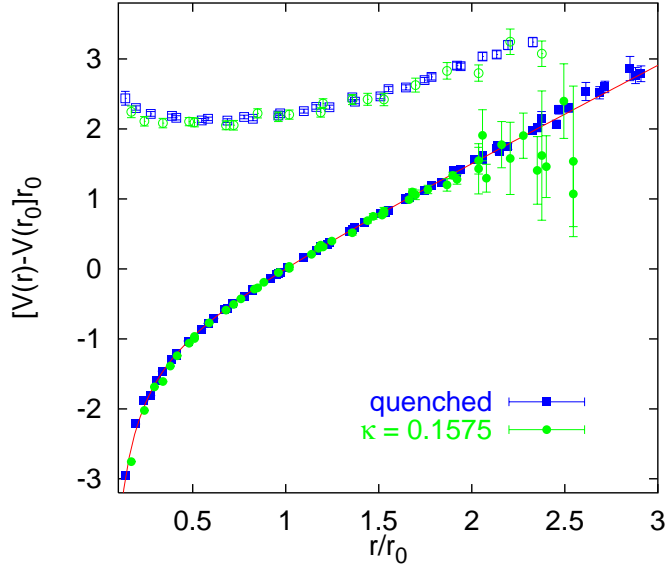}}\vspace{-.5cm}
  \caption[x]{Ground state (full symbols) and $E_u$ hybrid potential
(open symbols) at $\beta=6.2$ (quenched) and $\beta=5.6$, $\kappa=0.1575$
\cite{tcl}.}
  \label{fig:pot}
  \end{centering}\vspace{-.3cm}
\end{figure}

Expectation values of operators $O$ are
determined by computation of the path integral,
\begin{equation}
\langle O\rangle=\frac{1}{Z}\int[dU][d\psi][d\bar{\psi}]\,
O[U]e^{-S[U,\psi,\bar{\psi}]}.
\end{equation}
The normalisation factor $Z$ is such that $\langle 1\rangle =1$.
$U_{x,\mu}\in SU(N_c)$ denotes a gauge field and the
$4\times n_f\times N_c$ tuple $\psi_x$ is a fermion field.
The high-dimensional integral is evaluated by means of a 
(stochastic) Monte-Carlo method
as an average over an ensemble of $n$ {\em representative} gauge
configurations,
${\cal C}_i=\{U_{x,\mu}^{(i)}\}, i=1,\ldots,n$.
Therefore, the result on the
expectation value is subject to a statistical error
that will basically decrease like $1/\sqrt{n}$:
the longer we {\em measure} the more precise the prediction becomes.
Therefore, we might speak of {\em lattice measurements} and
{\em lattice experiments} in analogy to {\em ``real'' experiments}. Ideally, 
the sample size $n$ is such that the statistical precision is
similar or smaller than the systematic uncertainty of the
extrapolations.

In Fig.~\ref{fig:pot} I illustrate that confinement is a numerical fact by
displaying the potential between a static quark and an anti-quark, separated
by a distance $r$.
The data have been obtained in $SU(3)$
gauge theory with and without two light (mass-degenerate)
quark flavours at lattice spacings somewhat smaller than 0.1~fm. 
The scale $r_0$ corresponds to 0.5~fm.
The (partially) unquenched potential,
that has been obtained at a quark mass $m_q\approx 50$~MeV, exhibits
a more pronounced singularity at small $r$, due to the slower
running of the QCD coupling towards {\em zero} as the momentum
scale is increased (or $r$ is decreased). Furthermore,
from $r\approx 1.2$~fm onwards, the (partially) unquenched
potential eventually flattens: the
QCD string ``breaks''.
Indeed, many interesting
features of the
theory can be studied on the lattice
that are not at all accessible to {\em real} experiment.
As an example,
I have included a so-called $E_u$ (or continuum $\Pi_u$)
hybrid potential~\cite{michael,paton}
into the plot that
corresponds to the interaction energy of two static quark sources
for the case that gluons contribute one unit of angular momentum along the
interquark axis.

\section{(Abelian) photons and monopoles}
\label{sec:abel}

Soon after the advent of QCD,
't~Hooft and Mandelstam~\cite{hooft}
proposed the dual superconductor scenario of
confinement; the QCD vacuum is thought to behave analogously
to an electrodynamic superconductor
but with the r\^oles of electric and magnetic fields being interchanged:
a condensate of magnetic monopoles expels electric fields
from the vacuum. If one now puts electric charge and anti-charge
into this medium, the electric flux that forms between them
will be squeezed into a thin, eventually
string-like, Abrikosov-Nielsen-Oleson (ANO) vortex which
results in linear confinement.

In all quantum field theories in which confinement has been proven,
namely in compact $U(1)$ gauge theory in the Villain
formulation~\cite{bankskogut},
the Georgi-Glashow model~\cite{georgi} and SUSY Yang-Mills
theories~\cite{wittenseiberg}, this scenario is indeed realised.
Before one can apply this picture to QCD or $SU(N_c)$ chromodynamics
one has to identify the relevant dynamical variables: it is not straight
forward to generalise the electro-magnetic duality of a $U(1)$ gauge theory
to $SU(N_c)$ since gluons carry colour charges. How can one define
electric fields in a gauge invariant way?
What fields are dual to the electric fields? How can one
identify a monopole current? The next question
is that for the order parameter of the confinement-deconfinement
phase transition: what symmetry is
broken? One might also ask why this symmetry is broken --- a question that
has been answered by the BCS theory for conventional superconductors.
Finally, we are interested in an effective
low-energy theory which corresponds to the
Ginzburg-Landau theory of standard superconductors.

Let us address the first question. In the
Georgi-Glashow model, the $SO(3)$ gauge symmetry is broken down to
a $U(1)$ symmetry as the vacuum expectation value of
the Higgs field becomes finite. Within this effective
$U(1)$ theory, the standard electro-magnetic
duality is realised, resulting in confinement. Such a mechanism is not
provided by QCD but one can attempt to reduce the $SU(N_c)$ symmetry
to an Abelian symmetry {\em by hand}. In this spirit,
it has been proposed~\cite{hooft2}
to identify the monopoles in the $U(1)^{N_c-1}$ diagonal Cartan subgroup
of $SU(N_c)$ gauge theory after gauge fixing in respect to the
off-diagonal $SU(N_c)/U(1)^{N_c-1}$ degrees of freedom (Abelian projection).

In general,
a gauge transformation,
$\Omega_x\in SU(N_c)/U(1)^{N_c-1}$, can be found
that diagonalises an arbitrary
operator $X_x=f[U^{\Omega}]$ within the adjoint representation
of $SU(N_c)$
when applied to the link variables,
\begin{equation}
U_{x,\mu}^{\Omega}=\Omega_xU_{x,\mu}\Omega^{\dagger}_{x+a\hat{\mu}}.
\end{equation}
Subsequently, a coset decomposition of the gauge transformed links
is performed,
\begin{equation}
U_{x,\mu}^{\Omega}=C_{x,\mu}u_{x,\mu}\quad\mbox{with}\quad
C_{x,\mu}\in SU(N_c)/U(1)^{N_c-1},\quad
u_{x,\mu}\in U(1)^{N_c-1}.
\end{equation}
If we now apply a residual gauge transformation, $\omega_x\in U(1)^{N_c-1}$,
we find,
\begin{equation}
u_{x,\mu}\longrightarrow\omega_x u_{x,\mu}\omega^{\dagger}_{x+a\hat{\mu}},
\quad
C_{x,\mu}\longrightarrow\omega_x u_{x,\mu}\omega^{\dagger}_{x},
\end{equation}
i.e.\ $u_{x,\mu}$ transforms like a gauge field while
$C_{x,\mu}$ transform like matter fields. Therefore, we will refer to
the diagonal gluon field $u_{x,\mu}$ as ``photon'' field.
While the gauge fields of
electrodynamics in the non-compact continuum formulation
are free of singularities and, therefore, cannot form
magnetic monopole solutions, the gauge transformation $\Omega_x$ is
singular wherever two eigenvalues of $X_x$ coincide. Therefore,
the gauge field $u_{x,\mu}$ will in general contain
such monopoles.

After Abelian gauge fixing QCD can be regarded as a theory of interacting
photons, monopoles and matter fields (i.e.\ off-diagonal gluons and quarks).
One might assume that the off-diagonal
gluons do not affect long range interactions.
This conjecture is known as {\em Abelian dominance}~\cite{ezawa}.
In addition, {\em monopole dominance} of non-perturbative physics has been
proposed~\cite{suzukirev}.

The identification of photon fields and monopoles is a
gauge invariant process. However, the choice of
the operator $X_x$, that defines the Abelian projection, is ambiguous.
In his original work, `t~Hooft suggested that the dual superconductor scenario
would be realised in any Abelian projection. Indeed,
the expectation value of a monopole creation operator has been
found to be an order parameter
of the de-confinement phase transition in quite a few different
projections of $SU(2)$ chromodynamics~\cite{digiacomo}. As long as 
any $U(1)^{N_c-1}$ projection of the theory yields similar results,
one does not have to specify the mechanism that is thought to
break the $SU(N_c)$ gauge symmetry of QCD.
However, numerical simulations
suggest that Abelian and monopole dominance are not at all universal.
The most popular gauge projection applied and discussed is the
maximally Abelian projection (MAP)~\cite{hooft2}. One feature that is not
shared by almost all other projections that have been investigated 
on the lattice so far is that the (local)
MAP gauge condition gives rise to non-propagating ghost fields only,
guaranteeing renormalisability of the Abelian projected theory.
Another --- and possibly related --- fact is that both, Abelian
and monopole dominance have qualitatively been verified in this projection.

One should mention that the analogy to an ordinary superconductor
after Abelian projection is not complete. The electrons that form Cooper pairs
in BCS theory are all negatively charged. However, the monopoles
that are thought to condense in QCD can carry both,
negative and positive magnetic charges. Therefore, the composition of the
condensate is very different. The origin of the
interaction that results in an attractive force between monopoles and
anti-monopoles also differs from the periodic background potentials
of the BCS theory.

I will briefly explain how MAP is performed on the lattice for the
case of $SU(2)$ gauge theory. In the first step one maximises the
functional~\cite{kronfeld},
\begin{equation}
F(\Omega)=\sum_{x,\mu}\mbox{tr}\left(\tau_3 U^{\Omega}_{x,\mu}\tau_3
U^{\Omega \dagger}_{x,\mu}\right),
\end{equation}
by means of a gauge transformation $\Omega_x$. $\tau_3$ denotes
a Pauli-matrix\footnote{In principle, we can fix the gauge along any direction
within the $SU(2)$ group space. The resulting gauges
will only differ from each other by a global gauge transformation that
will not affect expectation values --- as long as we perform the subsequent
coset decomposition with respect to the same $U(1)$ subgroup.
Also note that if we forced an adjoint Higgs field with a $\delta$-like
potential into the
3-direction, the form of the interaction term
with the gauge fields would be identical to $F$.}.
After the maximisation, all link variables are {\em as diagonal as possible}.
The resulting gauge fields $U^{\Omega}_{x,\mu}=
\exp\left(i\sum_cA_{x,\mu}^c\tau_c/2\right)$
satisfy 't~Hooft's differential MA gauge fixing
condition~\cite{hooft2},
\begin{equation}
\left(\partial_\mu\pm iA^3_{x,\mu}\right)A^{\pm}_{x,\mu}=0,\quad
A^{\pm}_{x,\mu}=\frac{1}{\sqrt{2}}\left(A^1_{x,\mu}\pm A^2_{x,\mu}\right).
\end{equation}

After gauge fixing, the projection is performed:
observables are calculated on Abelian
configurations $\{\theta_{x,\mu}\}$
rather than $\{U_{x,\mu}\}$.
We refer to the field theory, defined in this way,
as Abelian projected $SU(2)$ gauge theory [APSU(2)].
For convenience the Abelian links are represented
in the Lie algebra rather than the group itself,
$u_{x,\mu}=\exp\left(i\theta_{x,\mu}\tau_3\right)$.
Note that the normalisation has been chosen such that the periodicity
of $\theta_{\mu}$ is $2\pi$ as opposed to the periodicity
$4\pi$ of $A^3_{\mu}$. With this convention we find magnetic charges
to be multiples of $2\pi/e$ just like in electrodynamics rather than
$2N_c\pi/e$ as one might have expected in $SU(N_c)$ gauge theory.
We also use the convention $e=1$ for the fundamental electric charge.

Recently, some articles have appeared whose authors claim to have
``proven'' Abelian dominance, either in general or specifically
in the MAP. In one case the argument is based on the fact that the
off-diagonal gluon fields acquire mass and, therefore, are thought
to affect ultra violet physics only, a statement that does not necessarily
apply to confining field theories. Other arguments are either based on
misinterpretations of the transfer matrix formalism or on rewriting
the original theory in terms of Abelian projected variables plus
perturbations (the off-diagonal fields) that are proportional
to a {\em small} parameter $\epsilon$. In order to avoid going too much
into details I only mention that both, the static potential
and Wilson loops in MAP are very different from
their counter parts in the original theory, even
for large distances. In case of the potential, the
asymptotic slope (string tension) comes out to be similar. However,
APSU(2) and $SU(2)$ potentials are shifted with respect to each other
by a substantial constant due to different self-energy contributions.
Any {\em ``proof''} that fails to account for the latter
fact is necessarily wrong to some extent.

\section{Results}
\label{sec:res}

\subsection{Some Definitions}
The language of differential forms~\cite{difforms}
turns out to be very convenient
for the present purpose and this is even more so on the lattice.
Let us start with a $U(1)$ 
field theory in $D=4$ dimensions. The anti-symmetric field 
strength tensor $F$ (a 2-form)
can be decomposed as follows,
\begin{equation}
F = dA + \delta\! *\!C,
\end{equation}
where $A$ denotes the standard magnetic four-potential (1-form) and
$C$ denotes an electric four-potential which is a 1-form on the
dual lattice. $A$ and $C$ are not uniquely determined but subject
to a $U_{el}(1)\times U_{mag}(1)$ gauge invariance since gradients
of scalar fields can be added. ``$d$'' denotes the exterior derivative
that, applied to an $n$-form, results in an $n+1$-form. ``$*$'' is the
pull-back operator that connects an $n$-form to a $(D-n)$-form on
the dual lattice while the dual derivative $\delta=*d*$ turns
$n$-forms into $(n-1)$-forms. We have $*^2=1$ and $d^2=\delta^2=0$.
We define a Laplacian $\Box=(-)^D(d\delta +\delta d)$.

In this notation, we obtain the generalised Maxwell equations,
\begin{equation}
\label{max1}
\delta F = j,\quad dF = *k.
\end{equation}
The electric current $j$ is a 1-form on the original lattice
while the magnetic (monopole)
current $k$ is a 1-form on the dual lattice.
In Landau gauge ($\delta A=0$ or $\delta C=0$, respectively)
this means,
\begin{equation}
j=\Box A,\quad k = \Box C.
\end{equation}

Note that, unlike the continuum four-potential $A$, the link angles
$\theta\in(-\pi,\pi]$
are subject to a $2\pi$ shift-periodicity. Therefore, 
the identification,
\begin{equation}
F=\frac{1}{a^2}\sin\left(d\theta\right)\left[1+{\cal O}(a^2)\right],
\end{equation}
is natural for electro-magnetic fields. 
We can factorise the
plaquette $d\theta\in(-4\pi,4\pi]$ into a regular part
$\overline{\theta}_{\Box}\in (-\pi,\pi]$ and a singular part
$m\in\{-2,-1,0,1,2\}$:
\begin{equation}
\label{fact}
d\theta =\overline{\theta}_{\Box}+2\pi m.
\end{equation}
While $dd\theta=0$, $d\overline{\theta}_{\Box}=-2\pi dm$ does not necessarily
vanish. Let us consider the magnetic flux through a (spatial) plaquette,
$\Phi_{mag}=a^2d\theta\left[1+{\cal O}(a^2)\right]$, which, in the absence of
monopole strings, would be identical to a contour integral of the vector
potential $A$ around the plaquette. However, in presence of monopoles,
only $\exp\left(i\int_{\Box}\!d^2\!f\,d\theta\right)=
\exp\left(i\int_{\partial\Box}\!dx\,A\right)$ holds: the argument can be
shifted
by multiples of $2\pi$ which, with the normalisation $e=1$, is the
elementary monopole charge. Thus,
$-2\pi m$ counts the monopole contribution to the magnetic flux through the
plaquette and $-2\pi dm$ corresponds to the ``flux'' out of a 
3-cube, such that~\cite{degrandtoussaint}
\begin{equation}
\label{monodef}
k=-2\pi *\!dm=*d\overline{\theta}_{\Box}
\end{equation}
is a magnetic monopole current\footnote{I assume the quantities to
be given in lattice units $a$.
$k$ has dimension $a^{-3}$, $F$ has dimension $a^{-2}$.}. $k$ is conserved
since $\delta k = -2\pi\delta^2 * m=0$, i.e.\ monopole currents
form closed loops on the dual lattice. The individual components of
$k$ can take values $2\pi n$ with $n=-4,-3,\ldots,4$.

Monopole currents can alternatively be defined through~\cite{faber},
$\tilde{k}=*dF=*d\sin\left(d\theta\right)$, the advantage being that
the second Maxwell equation [Eq.~(\ref{max1})] is automatically fulfilled.
The current $\tilde{k}$ is obviously conserved too.
The quantisation of magnetic charges, however,
is obscured by lattice artefacts. Magnetic monopoles become extended objects
and a geometric interpretation is
not as straight forward as for the definition presented before.
For all results I am going to review, the first definition has been
used.
Note that the locations of monopoles in either of the definitions
are not necessarily identical to positions of singularities of the
adjoint operator $X$ that has been diagonalised by the gauge fixing,
as originally proposed by 't~Hooft.

Each link variable can be factorised~\cite{bali,lee,sijs}
into a {\em singular} part
$\theta^{sing}$ that
is induced by magnetic monopoles and a {\em regular} (or photon)
part $\theta^{reg}=\theta-\theta^{sing}$ that obeys the
homogenous Maxwell equation $dF=d\sin(d\theta^{reg})=0$.
If we take the divergence of Eq.~(\ref{fact})
in Landau gauge ($\delta\theta=0$) we obtain,
\begin{equation}
2\pi\delta m = \delta d\theta = \Box \theta^{sing}.
\end{equation}
One solution of this equation is obviously,
\begin{equation}
\theta^{sing}_x=-2\pi\sum_yD_{xy}\delta m_y,
\end{equation}
where
$D_{xy}$ denotes the lattice Coulomb propagator in position space.

Instead of using the Abelian links $\{\theta_{x,\mu}\}$ one can
evaluate observables from the {\em monopole} and {\em photon}
parts $\{\theta^{sing}_{x,\mu}\}$
and $\{\theta^{reg}_{x,\mu}\}$, separately. We will refer to such expectation
values as {\em monopole} and
{\em photon contributions}, respectively.

\subsection{Successes}
I will present some facts in support of the superconductor
confinement scenario in the MAP:
approximate Abelian dominance of the static potential has been
verified~\cite{suzukirev}. In a recent study on a large lattice~\cite{bali},
the APSU(2) string tension has been confirmed to account for
$(92\pm 4)\%$ of the $SU(2)$ string tension.
It is an open question whether the agreement will
improve as the continuum limit is approached.

\begin{figure}
  \begin{centering}
\epsfysize=8truecm\centerline{\epsfbox{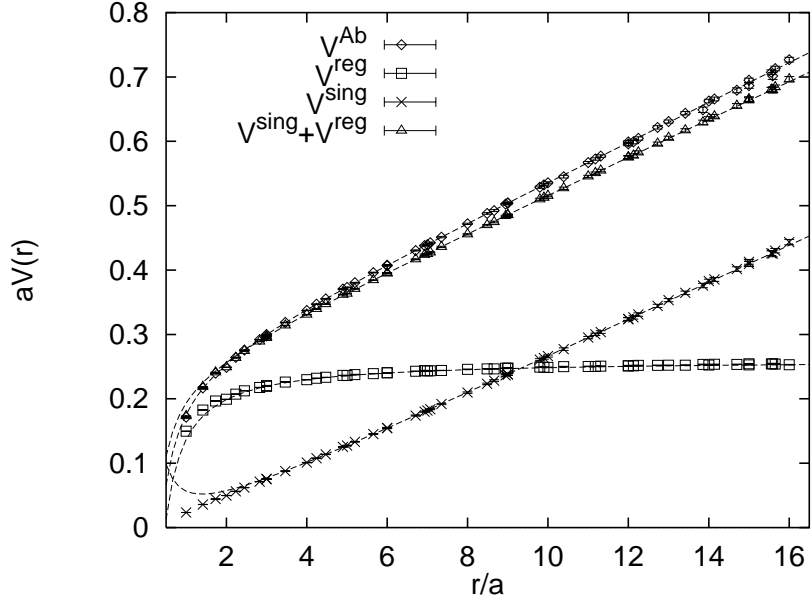}}\vspace{-.4cm}
  \caption[x]{Photon and monopole contributions $V^{reg}$ and
$V^{sing}$ to the APSU(2) potential~\cite{bali} $V^{Ab}$ in units
$a\approx 0.081$~fm.}
  \label{fig:monopot}
  \end{centering}\vspace{-.3cm}
\end{figure}

The photon part of the potential does not contain a linear
contribution.
Monopole dominance and the factorisation $V^{Ab}(r)=V^{sing}(r)+V^{reg}(r)$
approximately hold. In Fig.~\ref{fig:monopot},
we display the result of a recent lattice study~\cite{bali}.
Everything is plotted in lattice units $a\approx 0.081$~fm.
The monopole contribution amounts to $(95\pm 1)$\%
of the Abelian string tension.

\begin{figure}
  \begin{centering}
\epsfysize=8truecm\centerline{\epsfbox{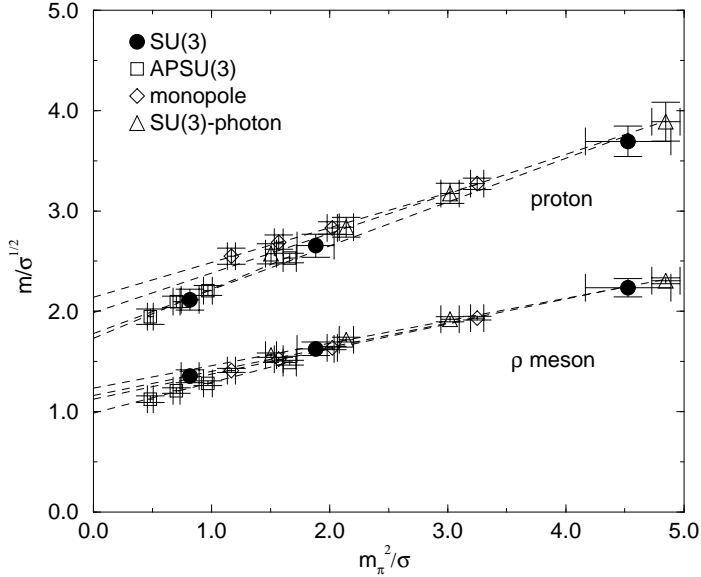}}\vspace{-.5cm}
  \caption[x]{$SU(3)$ $\rho$ and nucleon masses: Abelian,
monopole and non-photon contributions~\cite{suzuki2}.}
  \label{fig:hadron1}
  \end{centering}\vspace{-.3cm}
\end{figure}

Approximate Abelian and monopole dominance has been confirmed for the
light hadron spectrum~\cite{suzuki2} of $SU(3)$ gauge theory.
In Fig.~\ref{fig:hadron1}, I display recent results on the nucleon
and $\rho$ masses
that have been obtained on a lattice with extent $La\approx 2.8$~fm
and spacing $a\approx 0.175$~fm. On the horizontal axis,
the squared pion mass (in units of the string tension) is
displayed, which changes as the quark mass is varied.
Results for $SU(3)$, APSU(3) and Abelian monopole contributions
as well as for $SU(3)$ with the photon part removed lie on
the same curve. In the same study, $\pi$ and $\rho$ masses are found to
become degenerate, as soon as the monopole contribution to the gauge
fields is subtracted, both in $SU(3)$ chromodynamics and the
Abelian projected theory,
i.e.\ Abelian monopoles are required for chiral symmetry breaking.
Consistent results have been found by Lee {\em et al.}~\cite{lee}
and Bielefeld {\em et al.}~\cite{chiral}.
Of course, one can also argue that since instantons
are always accompanied by Abelian monopoles~\cite{bornyakov},
removing these monopoles naturally
results in a trivial vacuum topology.

The one-loop $\beta$-functions in APSU($N_c$) and $SU(N_c)$
gauge theories agree~\cite{reinhard}, i.e.\ at weak coupling, masses obtained
in both theories should be proportional to each other. A recent
analytical investigation has also confirmed anti-screening of
colour fields in APSU(2)~\cite{hart}. These results
are in agreement with numerical data~\cite{bali,inprep}.

For the next few observations, I refer to M.~Polikarpov's talk
at this conference: monopoles have been found to be condensed in the
confined phase of $SU(2)$ gauge theory~\cite{digiacomo,poli2}.
Furthermore, APSU(2) monopoles are spatially correlated with regions
of access in the $SU(2)$ action density~\cite{poli3} which rules out
that the monopoles are mere gauge fixing artefacts since their
presence is reflected in gauge invariant observables.
An effective monopole Lagrangian has been constructed which can
(approximately)
be mapped onto the Abelian Higgs model~\cite{poli}.

\subsection{Puzzles}

Of course, we do not expect $SU(N_c)$ gauge theories to be {\em identical}
to an Abelian Higgs model.
So, we should become suspicious if the
picture did not fail to describe the QCD reality
at some point. Obviously, APSU(2) becomes
very different from $SU(2)$ gauge theory in the ultra violet.
I will restrict myself to two points where something in the infra red
might go wrong that I consider as being serious.

The static potential between sources within the
adjoint representation of $SU(2)$ (adjoint potential) will
saturate at a distance at which the binding energy exceeds
the gluelump-gluelump threshold~\cite{michael}.
Therefore, we expect for the
adjoint string tension $\sigma_{adj}$ the asymptotic value,
\begin{equation}
\sigma_{adj} = \lim_{r\rightarrow\infty}\frac{dV_{adj}(r)}{dr}=0.
\end{equation}
In APSU(2) we obtain for adjoint Wilson loops around a rectangle
$S$,
\begin{equation}
W_{adj}^{Ab}(S)=\frac{1}{3}\left[1+2W^{Ab,2}(S)\right],
\quad W^{Ab,2}=\cos\left(2\!\!\!\sum_{(x,\mu)\in\partial S}\!\!
\theta_{x,\mu}\right),
\end{equation}
where we call $W^{Ab,2}$ the {\em charge two} Wilson loop; the
adjoint source corresponds to a {\em neutral} and two {\em charge two}
Abelian components.
Obviously, as the area $S$ goes to infinity, $W_{adj}^{Ab}(S)$
approaches the constant value $1/3$. Therefore,
\begin{equation}
V_{adj}^{Ab}=\frac{1}{a}
\lim_{t\rightarrow\infty}\log\frac{W(r,t)}{W(r,t+a)}
=0\quad\forall\quad r,
\end{equation}
i.e.\ the correct asymptotic value $\sigma_{adj}^{Ab}=0$ is produced.

\begin{figure}
  \begin{centering}
\epsfysize=8truecm\centerline{\epsfbox{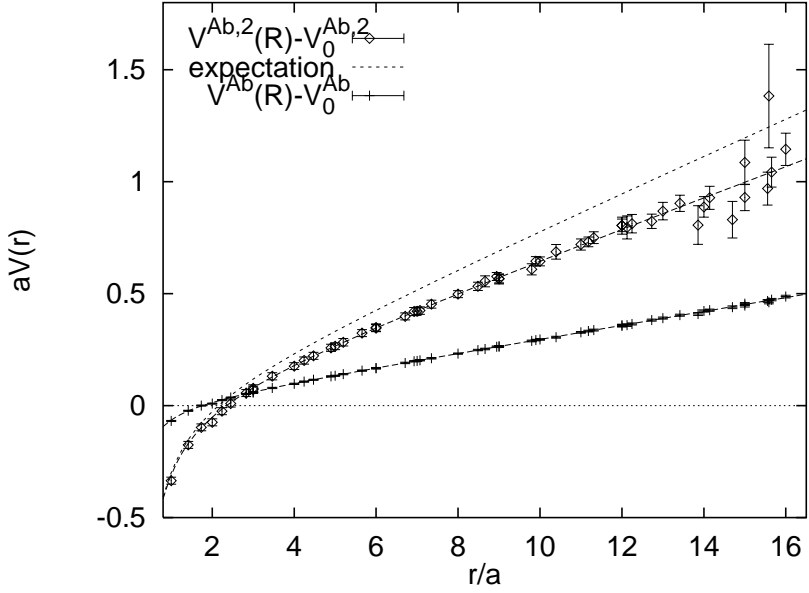}}\vspace{-.4cm}
  \caption[x]{The charge two potential in APSU(2) in lattice units
$a\approx 0.081$~fm~\cite{bali}.}
  \label{fig:charge2}
  \end{centering}\vspace{-.3cm}
\end{figure}

However, lattice results indicate that the adjoint string will
only break deep in the infra red at a distance slightly smaller
than 1.5~fm~\cite{michael}, a region in which
we would expect Abelian dominance to hold if we consider
the MAP theory to have relevance for
hadronic physics.
Simple models
predict a linear rise of the potential in the intermediate
region, with a slope that is
proportional to the Casimir charge of the representation, i.e.\
$\sigma_{adj}=8/3\,\sigma$ for $SU(2)$ gauge theory.
Indeed, lattice simulations yield a linear increase, however
with slightly smaller slope:
$2\sigma\leq\sigma_{adj}<8/3\,\sigma$.
In Fig.~\ref{fig:charge2} the fundamental potential $V^{Ab}(r)$
is displayed, together
with the {\em charge two} potential
$V^{Ab,2}(r)$~\cite{bali}. The
curves are fits in accord to the parametrisation,
\begin{equation}
V^{Ab(,2)}(r)=-\frac{e^{Ab(,2)}}{r}+\sigma^{Ab(,2)} r.
\end{equation}
The upmost curve corresponds to $e^{Ab,2}=4e^{Ab}$,
$\sigma^{Ab,2}=8/3\,\sigma^{Ab}$.
As in $SU(2)$, we find the slope in APSU(2) to be somewhat
smaller than expected.

In conclusion, it seems that the {\em charge two} potential resembles
the features of the {\em adjoint} potential within $SU(2)$.
However, it is not clear how we can get rid of the
{\em neutral} contribution to
the adjoint Wilson loop.
For this purpose, interactions with the
off-diagonal gluons have to be considered.

\begin{figure}
  \begin{centering}
\epsfysize=8truecm\centerline{\epsfbox{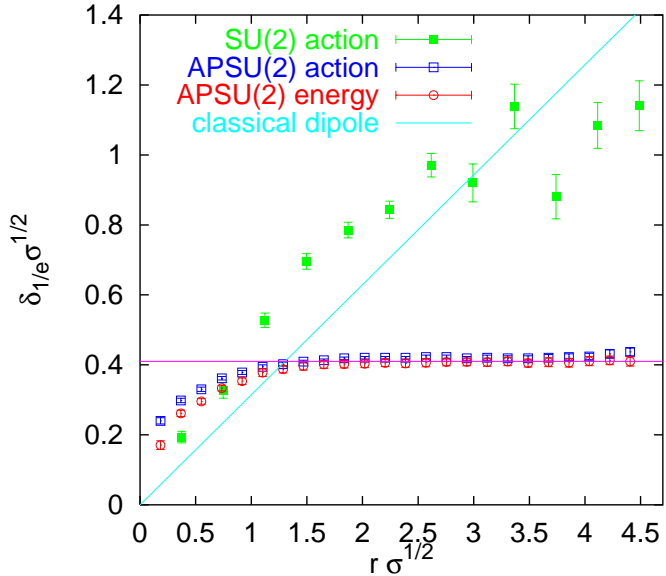}}\vspace{-.5cm}
  \caption[x]{$1/e$-radius of the flux tube in $SU(2)$ and
APSU(2)~\cite{inprep}.}
  \label{fig:width}
  \end{centering}\vspace{-.3cm}
\end{figure}

A second puzzle that has recently been noticed~\cite{inprep}
is related to the energy distribution within the
ANO vortex between static sources. In Fig.~\ref{fig:width}, I show
the $1/e$ radius of the flux tube as a function of the source separation.
The scale is set by $\sigma^{-1/2}\approx 0.45$~fm.
In $SU(2)$ gauge theory only for
the action density distribution a statistically reasonable signal
has been obtained while for
APSU(2) we are able to
display both, action and energy flux tube radii.
While the width of the action density distribution in $SU(2)$
increases before it eventually saturates around a separation
$r\approx 1.2$~fm at a value $2\delta_{1/e}\approx 0.9$~fm,
both APSU(2) widths become consistent with a constant
$2\delta_{1/e}\approx 0.36$~fm as soon as $r> 0.5$~fm.

Obviously, the matter fields change the structure of the
flux tube and, therefore, affect an infra red observable. However,
only the shape differs while the cross section,
i.e.\ the string tension, remains (almost) invariant. The tremendous width
of the $SU(2)$ flux tube is somewhat counter-intuitive in view
of the linear potential and the phenomenological success of string-like
descriptions of hadrons; obviously, the $SU(2)$ flux tube
is not one-dimensional
and transverse degrees of freedom should be considered. However,
the projected theory yields what we would have expected: a thin,
string-like object. We might speculate that the Abelian projection
reveals the physically relevant core of the vortex that
becomes obscured by the matter fields of the full theory.
However, we are left with a puzzle: why does the amplitude of
the underlying transverse string fluctuations remain
that constant? One should be able to reproduce such a vortex
in terms of an effective string theory. The simplest
such theories are not renormalisable
in {\em four} dimensions. Therefore, the width should increase
logarithmically with the rate being controlled by an ultra violet
cut-off, i.e.\ the scale at which longitudinal
fluctuations become important and the string description breaks
down. This is obviously not the case for our APSU(2)
``string''.
What is the physical mechanism that prevents the string from
widening? Is the effective string theory supersymmetric~\cite{greensite2}?
What correction terms have to be added in order to
produce such a string?

\subsection{The Dual Superconductor in Detail}
\label{sec:gl}

\begin{figure}
  \begin{centering}
\epsfysize=7.5truecm\centerline{\epsfbox{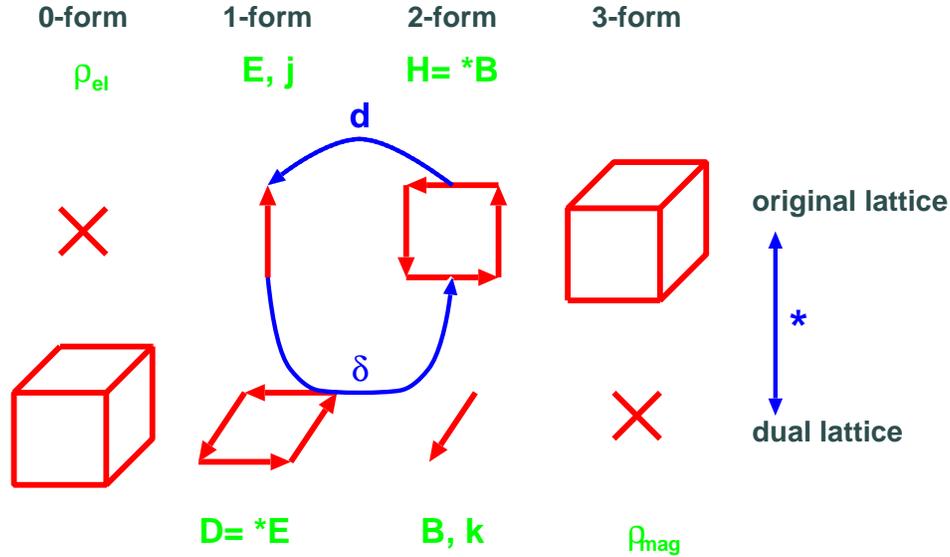}}\vspace{-.5cm}
  \caption[x]{Differential forms in $D=3$ dimensions.}
  \label{fig:forms}
  \end{centering}\vspace{-.3cm}
\end{figure}

In order to obtain an effective low-energy Lagrangian
with monopoles and photons as fundamental degrees
of freedom one can attempt to determine the free parameters
by numerically matching
the effective action to that of APSU(2)~\cite{poli,suzukirev}.
To complement such studies, one might probe the APSU(2) vacuum with electric
(or magnetic) test charges to verify predictions of the effective theory and
{\em measure} the values of the model parameters, which is the line
I am going to follow here.
Investigations of field distributions
in presence of charges have been performed previously~\cite{haymaker}. I 
will concentrate on the results from a more recent study~\cite{bali2}.

\begin{figure}
  \begin{centering}
\centerline{\epsfysize=5.5truecm\epsfbox{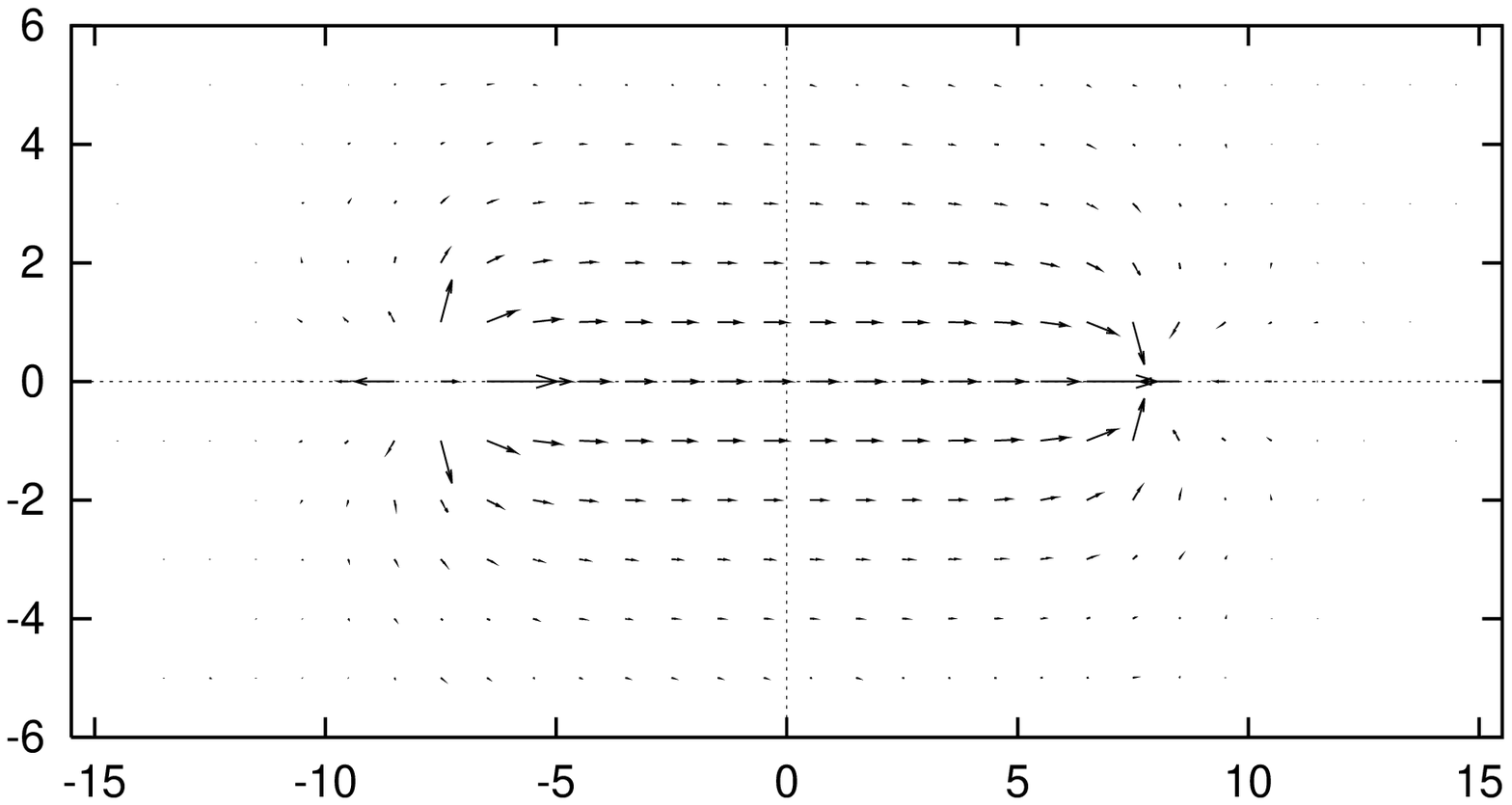}\hskip -.5cm\epsfysize=5.5truecm\epsfbox{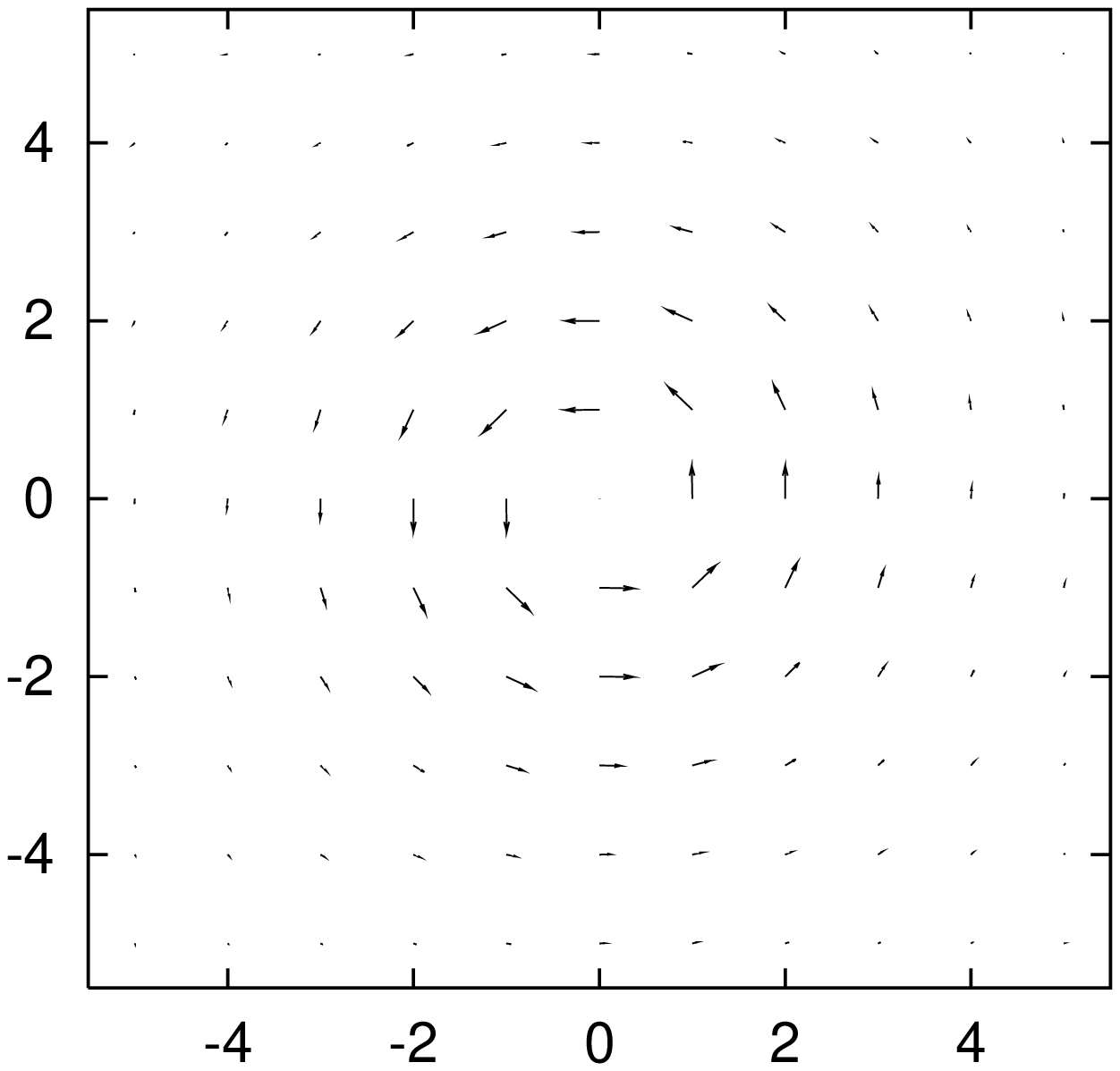}}\vspace{-.5cm}
  \caption[x]{Electric field ${\mathbf E}$
and magnetic super current ${\mathbf k}$ between two static sources.}
  \label{fig:E}
  \end{centering}\vspace{-.3cm}
\end{figure}

We are probing the vacuum with {\em static} electric sources. For this
purpose we consider three dimensional
spatial cross sections (time slices) of the lattice. In Fig.~\ref{fig:forms},
I have visualised where on the lattice different objects are ``living''.
The advantage in working with differential forms is that
the Stokes theorem is guaranteed to be exact and not subject
to lattice artefacts: if the differential Maxwell equations
are fulfilled, the integrated versions automatically hold too.
Of course, when finally relating
lattice results to continuum physics, lattice artefacts re-enter
the game.
The (generalised) Maxwell equations [Eq.~(\ref{max1})] read,
\begin{eqnarray}
\delta {\mathbf E} &=& \rho_{el},\quad d{\mathbf E} = *{\mathbf k}
-*\dot{{\mathbf B}},\\
\delta {\mathbf B} &=& \rho_{mag},\quad d{\mathbf B} = *{\mathbf j}
+*\dot{{\mathbf E}}.
\end{eqnarray}
The charge densities $\rho_{el}$ and $\rho_{mag}$ are the 4-components
of $j$ and $k$, respectively. The ``{\em dot}''-symbol denotes 
a temporal derivative.
For a static problem, electric and magnetic fields decouple.
In the absence of magnetic test charges, this implies
$\rho_{mag}=0$, ${\mathbf j}={\mathbf 0}$ and,
therefore, ${\mathbf B}={\mathbf 0}$
such that we are left with
two Maxwell equations only,
$\delta {\mathbf E} = \rho_{el}$ and $d{\mathbf E}=*{\mathbf k}$.

As mentioned above, one can define the monopole current via the
latter equation (Amp\`ere's law)
which is then trivially fulfilled. However, with
our definition of ${\mathbf k}$ [Eq.~(\ref{monodef})] neither of the
equations necessarily holds. Strong fields may give rise to non-linear
quantum corrections. Moreover, APSU(2) is not electrodynamics; in
general, the action will be non-local and the equations of motion might
differ. Our numerical study, however, verifies
$\mbox{div}\, {\mathbf E}$ to disappear outside
of the vicinity of the sources while the curl of the electric field
is identical to the magnetic
current~\cite{bali2}.
I display the electric field for
a distance $r=15a\approx 1.2$~fm in Fig.~\ref{fig:E}. The
vortex is stabilised by the surrounding super current ${\mathbf k}$.

The starting point of our investigation is the London limit.
The electric vector potential ${\mathbf C}$
on the dual lattice is defined through,
\begin{equation}
{\mathbf E}=*d{\mathbf C}.
\end{equation}
One representation of the (dual) London equations reads,
\begin{equation}
\label{eq:london}
{\mathbf C} + \lambda^2 {\mathbf k}=0,
\end{equation}
with $\lambda$ being the inverse mass of the dual photon.
By building the curl of this relation, we obtain
${\mathbf E}=-\lambda^2*d{\mathbf k}$ which, together with the dual
Amp\`ere law
results in,
\begin{equation}
\label{eq:superE}
{\mathbf E}=-\lambda^2\delta d{\mathbf E}=\lambda^2\left(
\nabla^2{\mathbf E}+d\rho_{el}\right).
\end{equation}
Without an electric current ${\mathbf j}$, the Maxwell equations
imply, $d\left({\mathbf B}+\lambda^2d\dot{\mathbf k}\right)=0$, i.e.\
even in the absence of a
magnetic field, the super current ${\mathbf k}$ remains
constant and in general non-{\em zero}.
{}From Eq.~({\ref{eq:superE}})
it is also obvious why
$\lambda$ is called the {\em penetration length}:
if we expose a superconducting probe to a homogeneous electric field,
${\mathbf E}=E_z{\mathbf e}_x$,
the field strength will decay
with the distance $x$ towards the centre of the medium:
$E_z(x)=E_z(0)\exp(-x/\lambda)$.

We create a charge-anticharge pair at a separation
$r$ parallel to the $z$-axis of our lattice,
$\rho_{el}=\delta^3(-r/2\,{\mathbf e}_z)-\delta^3(r/2\,{\mathbf e}_z)$.
$x$ denotes the transverse distance
from the core of the ANO vortex. For the electric flux
through any surface enclosing the charge, we expect $\Phi_{el}=
\int\!d^2x\,E_z(x,0)=1$.
For our geometry
Eq.~({\ref{eq:superE}}) reads in the centre plane perpendicular to
the vortex,
\begin{equation}
E_z(x)=\lambda^2\nabla_2^2E_z(x)+\Phi\delta^2(x),\quad
E_x(x)=0.
\end{equation}
A solution can be expressed in terms of the modified Bessel function $K_0$,
\begin{equation}
\label{eq:Lfit}
E_z(x)=\frac{\Phi_{el}}{2\pi\lambda^2}K_0(x/\lambda).
\end{equation}
Within statistical accuracy, we find our data to be compatible with such
a functional form for $x>x_{\min}=4.2a\approx 0.35$~fm\footnote{
Along off-axis lattice directions, we obtain data for non-integer
multiples of the lattice spacing.} with parameter values,
\begin{equation}
\lambda=(1.82\pm 0.07)a\approx(1.3\, \mbox{GeV})^{-1},\quad
\Phi_{el}=
1.44\pm 0.08.
\end{equation}
For small $x$ the data are overestimated by the fit since
$K_0(x)$ diverges as $x\rightarrow 0$ which explains why the electric
flux comes out to be significantly larger than {\em one}.
A dual photon mass of 1.3~GeV is compatible with
the mass of the lightest glueball in $SU(2)$ gauge theory.
However, the quantum numbers of a dual photon are $J^{PC}=1^{+-}$
as opposed to $0^+$ for this glueball.

How can we refine our description such that the electric field not
only in the surface region
but also closer to the centre of the vortex is correctly
reproduced? Obviously, a second scale $\xi$ that is
smaller than the $0.035$~fm, above which
the London limit seems to apply, has to be introduced.
Such a scale appears in the Ginzburg-Landau (GL) equations as
the {\em coherence length} of the GL wave function
$\psi({\mathbf x})$ that describes the spatial density
of superconducting monopole charges,
$n({\mathbf x})=|\psi({\mathbf x})|^2$.
We decompose $\psi$ into a phase and an amplitude $f$,
\begin{equation}
\psi({\mathbf x})=\psi_{\infty}f({\mathbf x})e^{i\theta({\mathbf x})},
\quad
f({\mathbf x})\stackrel{\scriptsize x\rightarrow\infty}{\longrightarrow} 1.
\end{equation}

$\xi$ characterises the decay of the monopole density towards the centre
of the vortex, where $f$ will vanish as superconductivity breaks down,
while $\lambda$ controls the penetration of the
vortex field into the surrounding
vacuum. The case $\xi=0$ corresponds to the London
limit. If we increase the width of the flux tube, the dia-electric energy
of the vortex is reduced by an amount roughly in proportion to $\lambda E^2$
while the amount of energy we have to pay for pushing the monopoles
further into the vacuum increases like $\xi E^2$. 
Values
$\xi<\sqrt{2}\lambda$ correspond to a negative surface energy in
accord to the Abrikosov criterium for a classical system.
This tendency to maximise the surface
results in retardation of flux tubes: we obtain a type II
superconductor while values $\kappa=\lambda/\xi<1/\sqrt{2}$ correspond
to a type I superconductor. From our experience we are prejudiced to
expect a type II scenario since in electrodynamics
type I flux tubes cannot be realised
due to the absence of isolated magnetic charges.
In the present situation, however, the presence
of two isolated electric sources
forces
the field lines through the surrounding vacuum, regardless of
the type of the superconductor.

If we restrict ourselves to the perpendicular centre plane, 
the equation $d{\mathbf C}=*{\mathbf E}$ implies for the
azimuthal component of ${\mathbf C}$
(up to gauge transformations),
\begin{equation}
\label{eq:integr}
C_{\theta}(x)=\frac{1}{2\pi x}\int_{x'<x}\!d^2x'\,E_z(x')\stackrel
{\scriptsize x\rightarrow\infty}{\longrightarrow}\frac{\Phi_{el}}{2\pi x},
\end{equation}
while the other components vanish. For non-constant
density of magnetic charges, the London equation Eq.~(\ref{eq:london})
is modified and becomes the second GL equation,
\begin{equation}
\label{eq:gl2}
\left(C_{\theta}(x)-\frac{\Phi_{el}}{2\pi x}\right)+\frac{\lambda^2}{f^2(x)}
k_{\theta}(x)=0.
\end{equation}
We can solve this equation with respect to $F(x)=f(x)/\lambda$ after
having reconstructed $C_{\theta}(x)$ via Eq.~(\ref{eq:integr}).

\begin{figure}
  \begin{centering}
\epsfysize=8truecm\centerline{\epsfbox{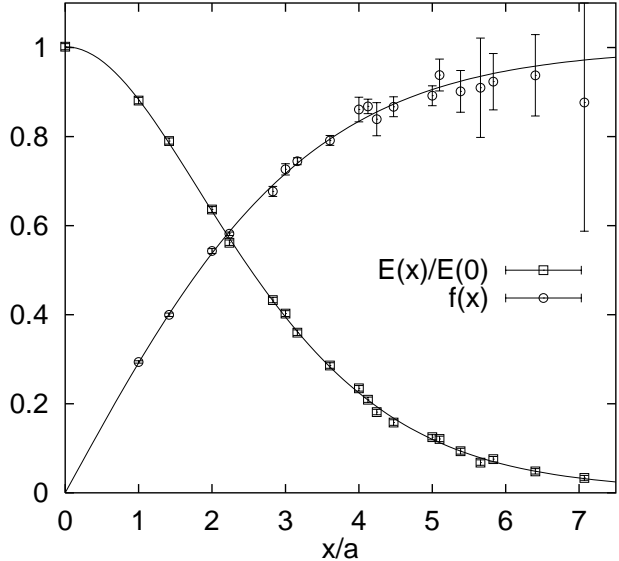}}\vspace{-.5cm}
  \caption[x]{The electric field and the amplitude of the
Ginzburg-Landau wave function against the distance from the centre of
the ANO vortex~\cite{bali2}.}
  \label{fig:wave}
  \end{centering}\vspace{-.3cm}
\end{figure}

The result
is displayed in Fig.~\ref{fig:wave}, together with $E_z(x)$. Data obtained at
$x< 2.2a$ has to be treated with care since the difference
between lattice and continuum versions of ``curl'' turns out to be
bigger than our statistical uncertainty. For $x>4.2a$
the errors on $f$ explode: here, no contradiction to the
London limit has been found.
We fit $F(x)$ with the ansatz,
\begin{equation}
\label{eq:Ffit}
F(x)=\frac{f(x)}{\lambda}=\frac{1}{\lambda}\tanh(x/\alpha),
\end{equation}
which conforms to the right boundary conditions. The fit is included
into the figure as well as the result of a fit of $E_z$ to a more involved
four parameter ansatz that also respects the boundary conditions
on $f$~\cite{bali2}.
From the fit Eq.~(\ref{eq:Ffit}) we obtain $\lambda=1.62(2)a$.
The fit to $E_z$ yields $\lambda=1.84(8)a$ while a 
simultaneous fit to $E_z$ and $k_{\theta}$ yields
$\lambda=1.99(5)a$. This has to be compared with the value
$\lambda=1.82(7)a$ from the London limit fit of Eq.~(\ref{eq:Lfit}).
We end up with the conservative estimate,
\begin{equation}
\lambda=1.84^{+20}_{-24}a=(0.15\pm 0.02)\, \mbox{fm},\quad
\Phi_{el}=1.08\pm 0.02.
\end{equation}
One should settle in a scaling study whether the
deviation from $\Phi_{el}=1$ can be attributed to a non-trivial
vacuum dielectricity constant due to anti-screening.

\begin{figure}
  \begin{centering}
\epsfysize=8truecm\centerline{\epsfbox{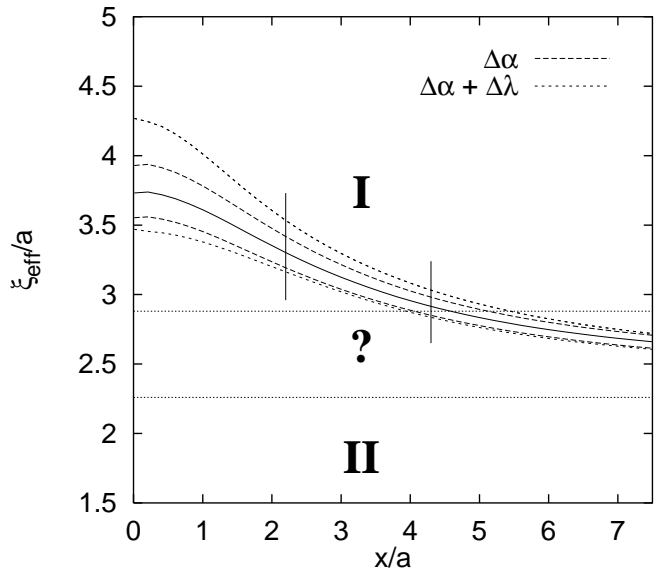}}\vspace{-.5cm}
\caption[x]{Effective coherence length versus distance from the centre
of the ANO vertex~\cite{bali2}.}
  \label{fig:type1}
  \end{centering}\vspace{-.3cm}
\end{figure}

Since the first GL equation is non-linear, we cannot consistently
formulate it in terms of differential forms on a lattice, i.e.\
--- unlike the Maxwell equations --- we have to
verify it in the {\em continuum} and discard
data obtained at small $x/a$ values where the lattice structure is
still apparent.
For our geometry the first GL equation reads,
\begin{equation}
f(x)=f^3(x)+\xi^2 h(x)
\left\{\left[\frac{1}{x}-\frac{2\pi C_{\theta}(x)}{\Phi_{el}}
\right]^2-\frac{1}{x}\frac{d}{dx}\left(x\frac{d}{dx}\right)\right\}f(x).
\end{equation}
Our strategy is to solve this equation with respect to an
{\em effective} coherence length $\xi_{\mbox{\scriptsize eff}}(x)$,
employing the above mentioned parametrisations
of $E_z(x)$ and $f(x)$ to interpolate the data.
The result is visualised in Fig.~\ref{fig:type1}.
Results outside of the
window of observation $2.2<x/a<4.2$ are unreliable,
for small $x$ due to lattice artefacts and for large $x$ due to
lacking precision data on $f$.
The figure contains error bands that are related
to the uncertainties in $\alpha$ as well as in $\lambda$.
Within the window, $\xi_{\mbox{\scriptsize eff}}$
varies by only 10~\%, i.e.\ the GL equation is
qualitatively satisfied. Taking this variation of $\xi$ with $x$
into account, we obtain the result,
\begin{equation}
\xi=3.10^{+43}_{-35}a= (0.251\pm 0.032)\,\mbox{fm}.
\end{equation}
The central value corresponds to
$\xi=\xi{\mbox{\scriptsize eff}}(\xi)$.

For the ratio of penetration and coherence lengths we determine,
\begin{equation}
\kappa = 0.59^{+13}_{-14}<\frac{1}{\sqrt{2}},
\end{equation}
i.e.\ we have evidence of a {\em weak} type I superconductor.
However, we are rather close to the Abrikosov limit whose
position might
differ from $1/\sqrt{2}$ due to quantum corrections.
In order to finally settle the question of the type of the superconductor,
interactions between two flux
tubes should be investigated.
This has been done in recent studies of
the confined phase of $U(1)$ gauge theory~\cite{zachfaberskala}
as well as in
three-dimensional $Z_2$ gauge theory~\cite{gliozzi}, the result being
attraction in both cases, i.e.\ type I superconductivity.

\section{Summary and Open Questions}

The lattice is an ideal tool to test ideas on the confinement mechanism.
Many infra red aspects of QCD are reproduced in the maximally Abelian
projection. After the projection only the monopole contribution to the
original gauge fields seems to be relevant for most low-energy properties.
The dual Maxwell equations have been verified in APSU(2) and
the fields are adequately described by the dual Ginzburg-Landau 
equations with the values $\lambda=0.15(2)$~fm and $\xi=0.25(3)$~fm
for penetration and coherence length, respectively. These values
correspond to a (dual) photon mass
$m_{\gamma}\approx 1.3$~GeV$\approx 3\sigma$
and a Higgs mass of $m_H\approx 0.8$~GeV$\approx 2\sigma$,
the ratio of which, $\kappa=\lambda/\xi=0.59(13)$, indicates
the vacuum of $SU(2)$ gauge theory to be a (weak) type I 
superconductor.
It is demanding to clarify
whether flux tubes in $SU(N_c)$ gauge theory as well as
in the Abelian projected theory attract or repel each other.

Electric flux tubes are found to be significantly thinner after the
Abelian projection. Contrary to $SU(2)$ gauge theory, their
width seems to saturate at a separation $r\approx 0.5$ fm
at a value, $2\delta_{1/e}\approx 0.36$~fm. The Abelian projection
also seems to suffer under problems with charges in non-fundamental
representations. These observations require further thought.
In the end we would like to clarify the r\^ole that charged
gluons play. Ideally, one would like to completely
circumvent the Abelian projection and arrive at a genuinely
non-Abelian description of the superconductor scenario.
In view of the existence of other reasonable proposals for the 
confinement mechanism more thought should be spend on relations
between different such pictures.

An extension of the detailed superconductor study presented
from $SU(2)$ to $SU(3)$ gauge theory should be attempted.
It is interesting to simulate the 4D Abelian Higgs model with the
$m_{\gamma}$ and $m_H$ parameters that have been predicted above and
compare the resulting flux distributions with the result
from APSU(2). This approach has recently been pursued by
Chernodub and collaborators~\cite{poli}.

\section*{Acknowledgements}

This work has been supported by
DFG grant~Ba~1564/3.
I thank my collaborators, in particular
V.~Bornyakov, M.~M\"uller-Preu\ss{}ker,
K.~Schilling, and C.~Schlichter.

\end{document}